\documentclass{nature}
\usepackage{graphicx}
\usepackage{amssymb}
\usepackage{float}
\usepackage{bm}
\usepackage{array}
\usepackage{amsmath}
\usepackage{booktabs} 
\usepackage{multicol}

\graphicspath{{Pic/}}

\title{Improved Earthquake Forecasting Model Based on Long-term Memory in Earthquake}

\author{Yongwen Zhang$^{1,2,3}$, Dong Zhou$^{4,1}$, Jingfang Fan$^{5,2}$, Warner Marzocchi$^{6}$, Yosef Ashkenazy$^{1}$ and Shlomo Havlin$^{2}$}

\begin{document}

\maketitle

\begin{affiliations}
 \item Department of Solar Energy and Environmental Physics, The Jacob Blaustein Institutes for Desert Research, Ben-Gurion University of the Negev, Midreshet Ben-Gurion 84990, Israel;
 \item Department of Physics, Bar-Ilan University, Ramat Gan 52900, Israel;
 \item Data Science Research Center, Faculty of Science, Kunming University of Science and Technology, Kunming 650500, Yunnan, China;
 \item School of Reliability and Systems Engineering, Beihang University, Beijing, 100191, China;
 \item Potsdam Institute for Climate Impact Research, 14412 Potsdam, Germany;
 \item Department of Earth, Environmental, and Resources Sciences, University of Naples, Federico II, Complesso di Monte Sant’Angelo, Via Cinthia, 21 80126 Napoli, Italy. 
 \end{affiliations}

\begin{abstract}
 A prominent feature of earthquakes is their empirical laws including memory (clustering) in time and space. Several earthquake forecasting models, like the Epidemic–Type Aftershock Sequence (ETAS) model\cite{Ogata1988, Ogata1998}, were developed based on earthquake empirical laws. 
 Yet, a recent study\cite{Zhang2019} showed that the ETAS model fails in reproducing significant long-term memory characteristics found in real earthquake catalogs. Here we modify and generalize the ETAS model to include short- and long-term triggering mechanisms, to account for the short- and long-time memory (exponents) recently discovered in the data. Our generalized ETAS model reproduces accurately the short- and long-term/distance memory observed in the Italian and South California earthquake catalogs.
 The revised ETAS model is also found to significantly improve earthquake forecasting.
\end{abstract}

The forecasting of the timing, location, and magnitude of future earthquakes is a long standing important problem. Past massive earthquakes resulted in catastrophic effects on human life, infrastructure and property and therefore finding early earthquake precursors is of great importance. It is accepted that even a small improvement in earthquake forecasting can save many human lives.
Currently, the understanding and forecasting of earthquake is limited. Predicting earthquakes by using a diagnostic precursor via some signal observable has not produced a reliable prediction scheme\cite{Jordan2006,Jordan2011,Ogata2017}. Actually, the current predictability of earthquakes is based on the known seismic laws: The distribution of earthquake magnitudes is exponential and follows the Gutenberg-Richter law ($N(m) \propto 10^{-bm}$, where $N$ is number of earthquakes of magnitude $m$ and $b \approx 1$)\cite{Gutenberg1944a}. The number of earthquakes triggered by a mainshock increases exponentially with the magnitude of the mainshock (Utsu law)\cite{UTSU1972}. In addition, the rate of triggered events decays as a power law with time (Omori law)\cite{Utsu1961}.

An operational earthquake forecasting scheme has been developed and applied to forecast earthquake sequences based on the empirical laws, including clustering of earthquakes in space and time\cite{Jordan2011}; space-time earthquake clustering can be attributed to triggering of earthquakes\cite{Zaliapin2008}. The space-time clustering of earthquakes stimulated the development of a series of earthquake forecasting models based on a branching process, such as the Epidemic--Type Aftershock Sequence model (ETAS)\cite{Ogata1988, Ogata1998}, and the Short--Term Earthquake Probability model (STEP)\cite{Gerstenberger}. The ETAS model
combines the Gutenberg-Richter, Utsu, and Omori laws into a Hawkes (point) process. In the ETAS model, every past earthquake (above a certain magnitude) triggers other earthquakes. Previous studies and many retrospective analyses\cite{Saichev2005,Woessner2011} proved that clustering models, such as the ETAS model, provide better forecasts compared to other models; still, these models lack some central earthquake features\cite{Vere-Jones2005,Lippiello2008,Gulia2018,Gulia}. Yet, it is possible that earthquake are erratic in their nature and, as such, are almost unpredictable. Indeed, many statistical physics features of seismic activity have not yet been fully understood\cite{DeArcangelis2016}.

Temporal and spatial memory (correlations) widely exist in many natural systems\cite{Koscielny-Bunde1998,Bunde2000} including in earthquake activity. For example, Livina et al.\cite{Livina2005} identified short-term memory of successive interevent times in real earthquake catalogs using a conditional probability method. They found a strong short-term memory where short (long) interevent time tends to follow short (long) interevent time. Other correlation detection methods like the detrended fluctuation analysis (DFA)\cite{Peng1994} have also been applied to detect the memory of interevent times\cite{Lennartz2008}. The empirical short-term memory between successive interevent times in real catalogs has been found to be reproduced by the ETAS model, only for a narrow range of model parameters\cite{Fan2018a}. Recently, a new measure has been\cite{Zhang2019} introduced, called ``lagged'' conditional probability, to explore long-term memory, both in successive and non-successive interevent times and distances\cite{Zhang2019}. This analysis resulted in a memory measure versus (time or distance) lag for which a crossover between two distinct behaviors has been found. A slow decaying power law for short scales (time or distance) while significantly faster decay (that might be exponential) at long scales has been found\cite{Zhang2019}. This behavior, discovered in real catalogs, could not be reproduced by the ETAS model. More specifically, the model's analysis resulted in single power law behavior (exponent) without a crossover that was observed in the real catalogs\cite{Zhang2019}; the model's memory is weaker (stronger) in short (long) time scales compared to the real catalogs.
The value of the power-law exponent depends on the productivity parameter $\alpha$, a parameter in the model which is associated with the Utsu law. Earthquakes can trigger more correlated events with a larger $\alpha$, resulting in enhanced earthquake memory. Therefore, based on the empirical finding\cite{Zhang2019} of crossover in the memory behavior, here we introduce into the ETAS model two productivity parameters, large and small, $\alpha_1$ and $\alpha_2$, for short and long time scales. We show here that this revised ETAS model reproduces the observed double power law behavior of memory, as well as the crossover observed in real data. Moreover, we show that the revised model improves significantly the forecasting performance of earthquake events.

\begin{figure}
\centering
\includegraphics[scale=0.6]{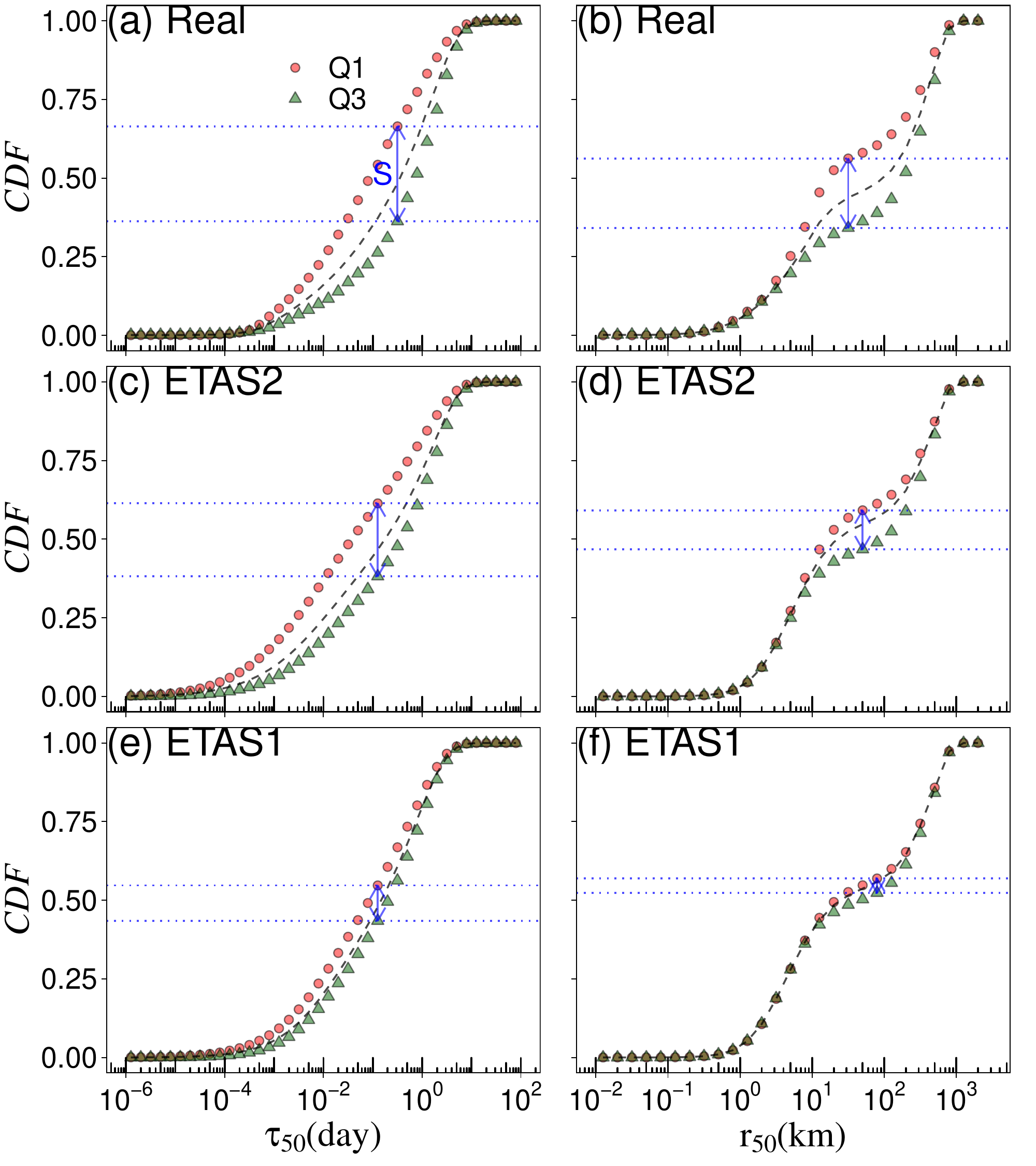}
\caption{\label{fig1}(Color online) Lagged conditional CDF of (a) interevent times $\tau_{k}$ and (b) interevent distances $r_{k}$ for lag-index $k=50$ (where the magnitude threshold is $M_{0} = 3.0$) for the real catalog of Italy; The maximum gap $S$ between the two lagged conditional CDFs (the first and third quantiles, $Q1$ and $Q3$), is indicated by the blue double arrow between the two dotted blue lines. The value of $0\leq S\leq1$ represents the memory.  (c), (d) Same as (a), (b) but for the (revised) ETAS2 model simulated catalog. (e), (f) Also same as (a), (b) but for the (original) ETAS1 model. The black dashed curves indicate the CDFs for all quantiles (all interevent series).}
\end{figure}

We first define the earthquake interevent time interval as $\tau_i=t_{i+1}-t_i$ (in days); this is the time interval between two consecutive earthquake events above a certain magnitude threshold. Similarly, an interevent distance $r_{i}$ is defined as the distance (in km) between the locations of events $i+1$ and $i$ above a certain magnitude threshold. We calculate the interevent times and distances with the magnitude threshold $M_0=3.0$ $(M_w)$ for the seismic catalog of Italy that is known to be complete for earthquake magnitudes above $3.0$ (see Methods); this catalog span 37 years, from 1981 to 2017. We then propose the ``lagged'' conditional CDF method based on the  ``lagged'' conditional PDF\cite{Zhang2019} as follows. First, all interevent times (distances) are sorted in ascending order and then divided into three equal quantiles. The first quantile, $Q1$, contains the smallest $1/3$ interevent times (distances) and the third quantile, $Q3$, contains the largest $1/3$ interevent times (distances). The conditional CDF of interevent times (distances) is defined as $C(\tau_k|\tau_0)$ ($C(r_k|r_0)$), where $\tau_0$ ($r_0$) belongs to $Q1$ or $Q3$, and $\tau_k$ ($r_k$) is the lagged $k$-th interevent time that follows $\tau_0$ ($r_0$). This method generalizes previous studies that used lag 1 ($k=1$)\cite{Livina2005}, thus enabling to uncover the empirical laws of long-range memory of earthquake catalogs. To demonstrate the empirical long term memory, we show in Figs. \ref{fig1}(a) and (b), the lagged conditional CDFs $C(\tau_{50}|\tau_0)$ and $C(r_{50}|r_0)$ using a high lag number $k=50$, for the first and third quantiles, $Q1$ and $Q3$, for the Italian catalog. It can be seen that the lagged conditional CDFs of $Q1$ are significantly different from that of $Q3$ (see Figure \ref{fig1}(a) and (b)). This implies the existence of memory (correlations) also for large number of lags. For the randomly shuffled catalogs which does not contain memory, both lagged conditional CDFs of $Q1$ and $Q3$ are identical to unconditional CDF (indicated by the black dashed curves in Figure \ref{fig1}(a) and (b)). Thus, the real catalogs exhibit memory, even for large $k$ lags, as found earlier\cite{Zhang2019}. 

We next test the memory in the ETAS model by simulating earthquakes records for the region of Italy ($34^\circ$N-$48^\circ$N, $6^\circ$E-$20^\circ$E), based on the thinning method\cite{Ogata1988, Ogata1998}. The original and our revised ETAS models are called here ETAS1 and ETAS2 respectively. In ETAS2, we introduce the productivity parameter $\alpha_1$ for small lag-index $k$ (short time scale), and the smaller productivity parameter $\alpha_2$ for large lag-index $k$ above a crossover value (long time scale). See the Methods section for more details regarding the model. Figure \ref{fig1}(c) and (d) show the lagged conditional CDFs $C(\tau_{50}|\tau_0)$ and $C(r_{50}|r_0)$ for the catalogs simulated by ETAS2. Note that there are substantial differences between the CDFs of $Q1$ and $Q3$, where these differences are common both for the real (Figure \ref{fig1}(a) and (b)) and for the ETAS2 simulated catalogs (Figure \ref{fig1}(c) and (d)). In contrast, the CDFs of Q1 and Q3 of the original ETAS1 model almost completely overlap, for both interevent times and distances (Figure \ref{fig1}(e) and (f)), in contrast to the observed CDFs. This demonstrates the existence of stronger memory in ETAS2 and, in contrast, much weaker memory in ETAS1.
                      
\begin{figure}
\centering
\includegraphics[scale=0.6]{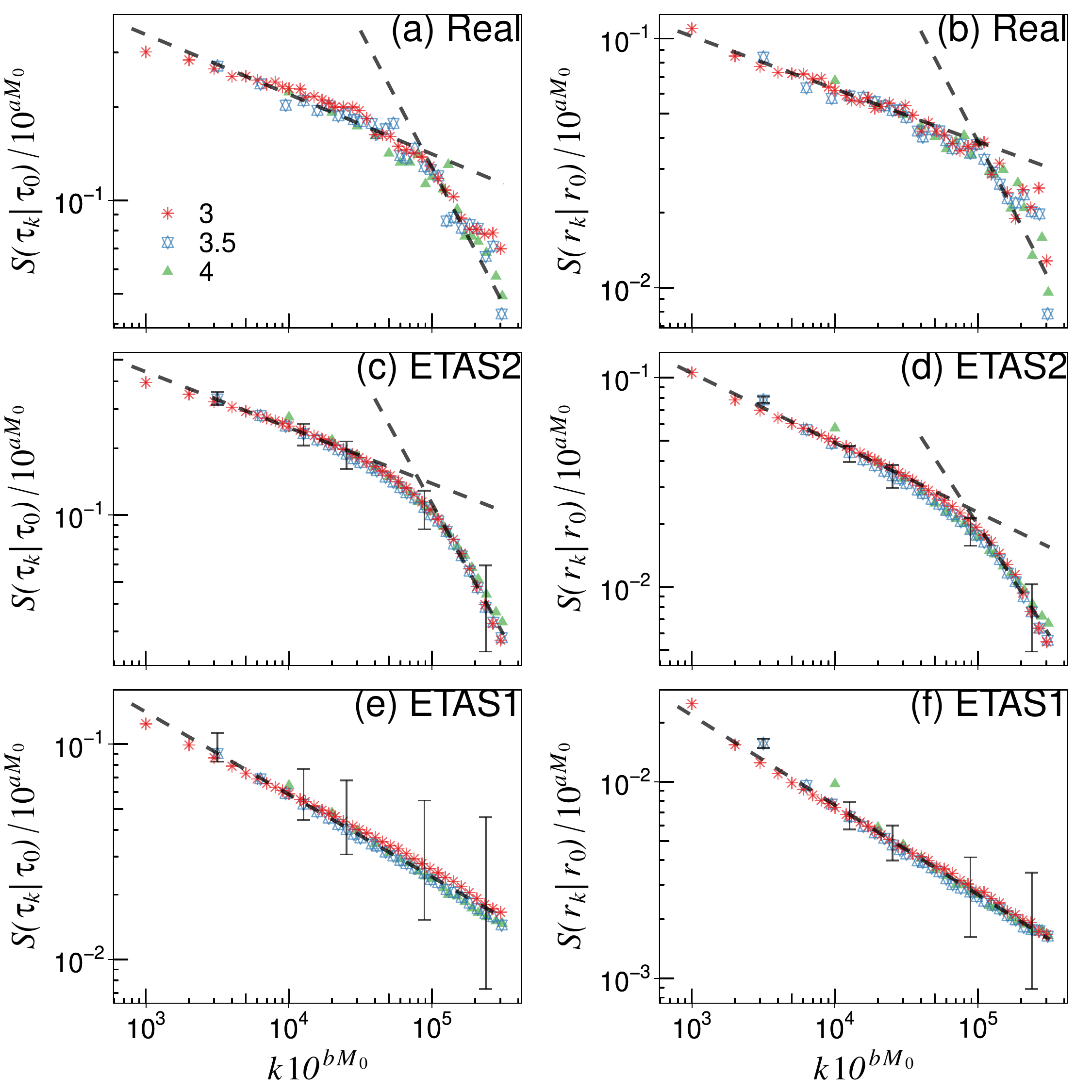}
\caption{\label{fig2}(Color online) The rescaled memory measure, $S$, (y axis) as a function of $k10^{bM_0}$ (x axis) for (a) interevent times and (b) interevent distances for the real Italian catalog. Colors and symbols represent different magnitude thresholds, $M_0$. (c), (d) Same as (a), (b) but for the simulated catalog of the new developed ETAS2 model. (e), (f) Same as (a), (b) but for the original ETAS1 model. The memory measure $S$ is averaged over $50$ independent realizations and the error bars show the $90\%$ confidence intervals. Black dashed lines are fitted power-law curves. The values of the exponents are given in Table \ref{T1}.}
\end{figure}

To quantify the memory based on the lagged conditional CDF, the maximum gap $S$ (indicated by double arrow in Figure \ref{fig1}) between the lagged conditional CDFs of the first and third quantiles is calculated. It follows that $S=0$ (complete overlap between Q1 and Q3 CDFs (PDFs)) indicates lack of memory while $S=1$ (complete separation between Q1 and Q3 PDFs) indicates full memory; thus, the memory measure is defined as $S(\tau_k|\tau_0)$ in the range between $0$ and $1$. We next calculate the memory measure for the real Italian catalog as a function of the lag index $k$ for interevent times and distances respectively (Supplementary Fig. 1). Different magnitude thresholds $M_0$ are also considered. We rescale the memory measure $S(\tau_k|\tau_0)$ by a factor $10^{aM_0}$ which represents the dependence of memory on the magnitude threshold\cite{Zhang2019} (i.e., $F(x)=S(x)10^{aM_0}$) where the lag-index, $k$, is rescaled by $x=k10^{bM_0}$, to account for the Gutenberg-Richter law. The curves $S(x)$ of all cases collapse into a single curve $F(x)$ after the rescaling (Figure \ref{fig2}). It is seen that the rescaled memory measure $F(x)$ of the Italian catalog decays slowly for small $x$ (lags) and faster for large $x$ (lags), both for interevent times and interevent distances, as seen in Figure \ref{fig2}(a) and (b). Figure \ref{fig2} (c) and (d) depict the corresponding scaling measures for the new ETAS2 model and it is clear that it reproduces the two power law behavior of the real catalogs (Figure \ref{fig2} (a), (b)) and the associated crossover of the scaling curves. Moreover, the crossovers are close to $x_c\approx 10^{5.0}$, for both the real catalog and our developed ETAS2 model. The averaged time corresponding to the crossover is around $100$ days. The scaling function behaves as $F(x) \sim x^{-\gamma_1}$ for $x<x_c$ and $F(x) \sim x^{-\gamma_2}$ for $x\geq x_c$. These two exponents ($\gamma_1$, $\gamma_2$) and the parameter $a$ are summarized in Table \ref{T1} showing that the new ETAS2 model reproduces quite accurately the scaling exponents of the real catalog. In contrast, the original ETAS1 model basically exhibits power-law behavior but with only a single exponent, without a crossover and thus fails in reproducing the observed memory characteristics of the data (see Figure \ref{fig2}(e) and (f)). We also calculate the memory measure for the real and simulated catalogs of South California (see Supplementary Figures 2, 3 and Table 1). The results indicate that the new ETAS2 model reproduces quite well the scaling function of the real catalogs, in contrast to ETAS1 model.
           

\begin{table}
\caption{\label{T1}%
Estimated parameters (mean $\pm$ std), $a$ exponent in the scaling factor $10^{aM_0}$ and power law exponents of the scaling function, $\gamma_1$, $\gamma_2$ for the interevent times and distances for the real catalog of Italy, ETAS1 and ETAS2 models.}
\centering
\begin{tabular}{ccccc}
\toprule
\textrm{}&
\textrm{Parameters}&
\textrm{Real}&
\multicolumn{1}{c}{\textrm{ETAS1}}&
\textrm{ETAS2}\\
\midrule
 & $a$ & 0.09 $\pm$ 0.01 & 0.17 $\pm$ 0.02 & 0.07 $\pm$ 0.01\\
Time & $\gamma_1$ &0.19 $\pm$ 0.01 &0.38 $\pm$ 0.01 &0.25 $\pm$ 0.01\\
& $\gamma_2$ &0.88 $\pm$ 0.05 &0.38 $\pm$ 0.01 &1.11$\pm$ 0.04\\
\midrule
 & $a$ & 0.24 $\pm$ 0.02 & 0.37 $\pm$ 0.03 & 0.21 $\pm$ 0.02\\
 Distance & $\gamma_1$ &0.21  $\pm$ 0.01 &0.46 $\pm$ 0.01 &0.33 $\pm$ 0.01\\
 & $\gamma_2$ &1.11 $\pm$ 0.07 &0.46 $\pm$ 0.01 &1.05 $\pm$ 0.03\\
 \bottomrule
\end{tabular}
\end{table}

\begin{figure}
\centering
\includegraphics[scale=0.6]{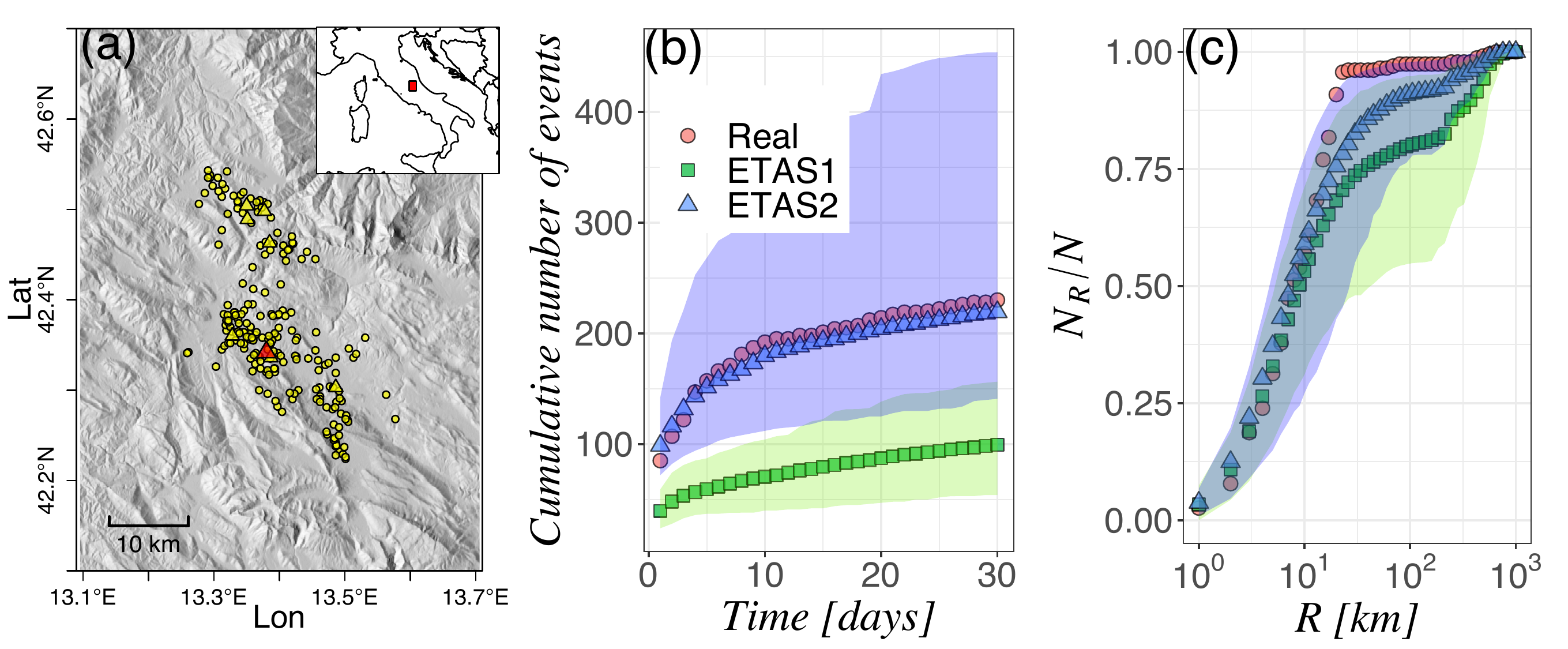}
\caption{\label{fig3}(Color online) Forecasting test for the L'Aquila earthquake sequences. (a) Locations of the L'Aquila earthquake sequences within $30$ days after the mainshock. The red triangle represents the mainshock ($M_w=6.3$), and the yellow triangles and dots represent the events after the mainshock (magnitude above $M_0=3.0$). Triangles indicate earthquake with magnitude above $5.0$. (b) Cumulative number of earthquakes as a function of days after the L'Aquila mainshock. (c) Fraction of the earthquakes within a radius $R$ centered at the mainshock epicenter, within $30$ days after the L'Aquila mainshock. The models' mean are averaged over $500$ independent realizations where the green and blue shadow areas indicate the $90\%$ confidence intervals for ETAS1 and ETAS2 models.}
\end{figure}

Next we test and compare the forecasting performance of ETAS1 and ETAS2 models. We perform the N-test\cite{Schorlemmer2007}, which compares the total number of earthquakes forecasted by the model with the total observed number of earthquakes over the entire region; we apply this test to the L'Aquila (Italy) earthquake (magnitude $6.3$), occurred on April 6, 2009\cite{Marzocchi2009}. Figure \ref{fig3}(a) presents the locations of earthquakes above magnitude threshold 3 occurred within one month after the L'Aquila mainshock. The number of earthquakes increased immediately after the L'Aquila mainshock (red circles in Figure \ref{fig3}(b)). Notably, the original ETAS1 model forecast much less events compared to the real catalog while the forecast of the revised ETAS2 is very similar to the real events. Indeed, the fact that the ETAS1 model severely underestimates the number of earthquakes immediately after large shocks has been reported previously\cite{Marzocchi2017}. The remedy was to specifically increase the $\alpha$-value after large shocks; still, this increase of $\alpha$-value is not consistent with the $\alpha$-value that was estimated based in the entire Italian catalog\cite{Marzocchi2009,Lombardi2010}. In the ETAS2 model that we propose here, we introduce in the model a large $\alpha_1$ in short time scale (below the crossover) such that the rate of aftershock is effectively increased immediately after a large shock; on the other hand, a small $\alpha_2$ is assumed for long time scales (above the crossover) that guarantees the averaged aftershock rate for long time period which is consistent with the rate of the entire real catalog. 

\begin{figure}
\centering
\includegraphics[scale=0.5]{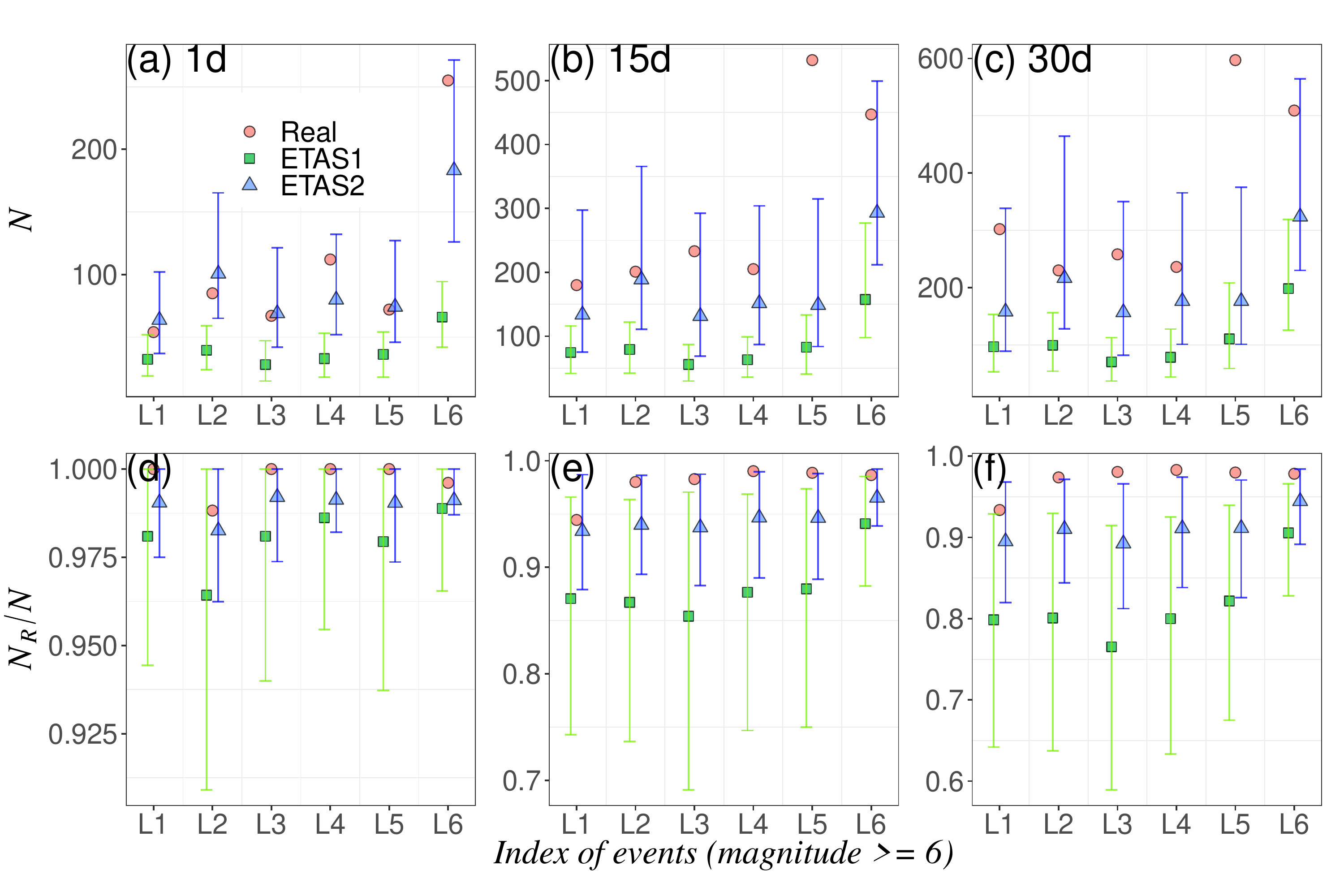}
\caption{\label{fig4}(Color online) Forecasting test after the six largest shocks of M$_w\geq6$ in Italy occurred between 1981 and 2017: L1 (1997 Umbria and Marche), L2 (2009  L'Aquila), L3 (2012 Emilia) and L4--6 (2016 Amatrice--Norcia). Total number of events after the mainshock within (a) $1$ day, (b) $15$ days and (c) $30$ days after the mainshock. Fraction of the earthquakes within a radius $100$ km from the mainshock epicenter within (d) $1$ day, (e) $15$ days and (f) $30$ days after the mainshock. The ETAS1 and ETAS2 model results are presented as averages over $500$ realizations where the error bars indicate the $90\%$ confidence intervals.}
\end{figure}

We also test the quality of forecast in space (Figure \ref{fig3}(c)) and find that our revised ETAS2 model outperform the original ETAS1 and it resulted in increased number of events, making its performance closer to real observations. Spatial clustering of aftershocks can cause a large fraction of events in short distance. Aftershocks are generally within a region of a radius $100$ km from the epicenter of the mainshock. We find that $98$\% of the events in Italy within $30$ days after the L'Aquila mainshock are within a radius of $R=100$ km (Fig. \ref{fig4}(c), (f)). We find that $94\%$ and $86\%$ events (from $500$ independent realizations) of ETAS1 and ETAS2 models occur within a radius of 100 km within one month from the main shock.


We also test the forecasts after six largest shocks (M$_w\geq6$) occurred in Italy between 1981-2017. Figs. \ref{fig4}(a), (b) and (c) show the results for the total number $N$ of earthquakes after the mainshock for $1$ day, $15$ days and $30$ days respectively. Almost for all these 6 major earthquakes the observed number of earthquakes (red dots) falls within the $90\%$ confidence intervals for new ETAS2 model forecasting. The exception is event L5 (Figs. \ref{fig4}(b), (c)), which occurred around the city of Norcia on October 26, 2016. This is because even a larger earthquake (L6, magnitude 6.6 in comparison to magnitude of 6.1 of the L5 event) hit Norcia later, leading to more aftershocks four days after L5. We also show in Fig. \ref{fig4}(d)--(f) the fraction of earthquakes within a radius $100$ km around the mainshock epicenter and within different number of days from the mainshock. The results indicate that the observed fraction of number of earthquakes (red dots) falls within the narrower error bars of the proposed ETAS2 model while the observations fall outside the error bars of the ETAS1 model, despite the larger error bars of this model (Fig. \ref{fig4}(d)--(f)). 
We thus conclude that the new ETAS2 forecasting is significantly better than the forecasting performance of ETAS1 model. Similar improved forecasting performances of ETAS2 after large shocks in South California are shown in Supplementary Figure 4.            

The spatiotemporal clustering of aftershock sequences dominates the memory in interevent times and distances. This is incorporated in the ETAS model. Our results suggest that actually the memory of the ETAS model depends on the aftershock productivity parameter $\alpha$. Following the observed catalogs\cite{Zhang2019} we revised the ETAS model to include two productivity parameters $\alpha_1$ and $\alpha_2$ for short and long time scales. We show that the revised ETAS model suggested above not only reproduces the memory characteristics (scaling function) of the real catalog
but also exhibits significantly better forecasting skills in comparison to
the original ETAS model. According to the Utsu law, the rate of aftershocks depends on the magnitude of the mainshock. The two $\alpha$--values of the revised ETAS model indicate that the Utsu relation for long-term triggering is not same as short-term triggering. In short time, a large earthquake tends to trigger much more events. After a characteristic ``time'', the triggering ability is substantially reduced. This may imply a possible mechanism regarding the triggering of aftershocks which depends on the stress conditions set up by historical events, not just a mainshock\cite{Parsons2014,Lippiello2015}. In other words, the aftershock triggering conditions could be changed by the sequentially dependent events occurred after the mainshock.

\section*{Methods}

\subsection{Data}
The Italian earthquake catalog is provided by the Seismic Hazard Center at Istituto Nazionale di Geofisica e Vulcanologia (INGV, http://terremoti.ingv.it/it/). This catalog spans the time period from 1981 to 2017 and it is complete for earthquake magnitudes above $M_0=3.0$; a catalog is ``complete'' when all events above the specified magnitude are included in the catalog. The South California catalog (https://scedc.caltech.edu/research-tools/alt-2011-dd-hauksson-yang-shearer.html) is from 1981 to 2018 and it is complete for earthquake magnitudes above $M_0=2.0$\cite{Hauksson2012}.

\subsection{The revised ETAS Model}
Seismic events are assumed  in the ETAS model to involve a space-time stochastic point process\cite{Ogata1988, Ogata1998}. Each event above a magnitude $M_0$ is selected independently from the Gutenberg--Richter distribution (where $b=1$). The conditional rate, $\lambda$, at location $(x, y)$ at time $t$  is given by
\begin{equation}\label{eq1}
\lambda\left(x, y, t | H_t\right)= \mu(x, y) + \sum_{t_i<t}k\left(M_i\right)g\left(t-t_i\right)f\left(x-x_i, y-y_i, M_{i}\right)\;,
\end{equation}
where $H_t$ is the history process prior $t$, $t_i$ are the times of the past events and $M_{i}$ are their magnitudes. $\mu(x, y)=\mu_{0}u(x, y)$ is the background intensity at location $(x, y)$, where $u$ is the spatial PDF of background events, which is estimated by the method proposed by Zhuang\cite{Zhuang2012}; $\mu_{0}$ is the background rate of the entire region. We represent the total number of the past events as $n-1$. The dependence of the triggering ability on magnitude is given by the Utsu law as     
\begin{equation}
k\left(M_i\right)=
\begin{cases}
A\exp(\alpha_1(M_i-M_0)) &  i > n-h10^{-bM_0}\\
A\exp(\alpha_2(M_i-M_0)) &  i \leq n-h10^{-bM_0}\;
\end{cases},
\label{eq2}
\end{equation}
where $A$ is the occurrence rate of earthquakes at zero lag. In Eq. \ref{eq2} we introduce two productivity parameters $\alpha_1$ and $\alpha_2$ ($\alpha_1\geq \alpha_2$). When $\alpha_1=\alpha_2$, Eq. \ref{eq2} reduces to the original ETAS model. $h10^{-bM_0}$ is the crossover number of events with the magnitude threshold $M_0$; $h$ is a parameter that can be estimated from the data. The $i$-th historical event has a larger rate to trigger the $n$-th event (due to the larger $\alpha_1$), when the number of events between the $i$-th and $n$-th events is smaller than $h10^{-bM_0}$.  

The function $g\left(t-t_i\right)$ follows the Omori law as      
\begin{equation}\label{eq3}
g\left(t-t_i\right)=\left(1+\frac{t-t_i}{c}\right)^{-p}\;,
\end{equation}
where $c$ and $p$ are the Omori law parameters. Spatial clustering of aftershocks is implemented by introducing a spatial kernel function $f\left(x-x_i, y-y_i, M_{i}\right)$\cite{Zhuang2012} as
\begin{equation}\label{eq4}
f\left(x-x_i, y-y_i, M_{i}\right)=\frac{q-1}{\pi \zeta^2}\left(1+\frac{(x-x_i)^{2}+(y-y_i)^{2}}{\zeta^2}\right)^{-q}, 
\end{equation}
where $\zeta=D\exp \left[\gamma_m\left(M_{i}-M_{0}\right)\right]$ indicates that the distances between triggering and triggered events depend on the magnitudes of triggering events. $q$, $D$ and $\gamma_m$ are the estimated parameters. 

The parameters of the original ETAS model (ETAS1) for Italy, were chosen as follows: $\mu=0.2$, $A=6.26$, $\alpha_1=\alpha_2=1.5$, $p=1.13$, $c=0.007$ based on refs. \cite{Fan2018a, Lombardi2015}. The spatial parameters were chosen to be $q=2.0$, $D=0.03$ (in units of ``degree'') and $\gamma_m=0.48$, and were estimated based on the method described in ref. \cite{Zhuang2012}. The parameters of the revised ETAS model (ETAS2) were chosen to be $A=3.26$, $\alpha_1=2.0$, $\alpha_2=1.4$ and $h=2\times10^5$ (the crossover $h10^{-bM_0}=200$ for the magnitude threshold $M_0=3.0$); the other parameters are the same as for the ETAS1 model.




\begin{addendum}
 \item We thank the Italian Ministry of Foreign Affairs and International Cooperation jointly with the Israeli Ministry of Science, Technology, and Space (MOST); the Israel Science Foundation, ONR, the Japan Science Foundation with MOST, BSF-NSF, ARO, the EU H2020 project RISE, and DTRA (Grants no. HDTRA-1-14-1-0017 and HDTRA-1-19-1-0016) for financial support. 
 \item[Competing Interests] The authors declare that they have no
competing financial interests.
 \item[Correspondence] Correspondence and requests for materials
should be addressed to Yongwen Zhang (email: zhangyongwen77@gmail.com).
\end{addendum}


\end{document}